\documentstyle[12pt,epsf]{article}
\def\qed{\nobreak\kern 1em \vrule height .5em width .5em depth 0em}
\def\vbar{\mathchoice{\vrule height6.3ptdepth-.5ptwidth.8pt\kern-.8pt}
   {\vrule height6.3ptdepth-.5ptwidth.8pt\kern-.8pt}
   {\vrule height4.1ptdepth-.35ptwidth.6pt\kern-.6pt}
   {\vrule height3.1ptdepth-.25ptwidth.5pt\kern-.5pt}}

\def\date
   {\noindent Date: \today \par
    \medskip}

\def\bra#1{\langle #1 |} 
\def\ket#1{| #1 \rangle}
\newcount\subno      
\newcount\subsubno   
\newcount\appno      
\newcount\appeqno    
\newcount\tableno    
\newcount\figureno   
\def\nsection#1
   {\vskip0pt plus.1\vsize\penalty-250
    \vskip0pt plus-.1\vsize\vskip24pt plus12pt minus6pt
    \subno=0 \subsubno=0
    \addtocounter{section}{1}
\renewcommand{\thesection}{\Roman{section}}
  {\small  \noindent {\bf \thesection. #1\par}}
    \noindent}
\def\nsubsecnn#1
   {\vskip-\lastskip
    \vskip12pt plus12pt minus6pt
    {\noindent {\bf #1\par}}
    \noindent}
\begin{document}
\setcounter{page}{1}
\pagestyle{plain}
\setcounter{equation}{0}
%
%
\ \\[12mm]
\begin{center}
    {\bf EXACT SOLUTION OF TWO-SPECIES BALLISTIC ANNIHILATION\\}
    {\bf WITH GENERAL PAIR-REACTION PROBABILITY}
\\[15mm]
\end{center}
\begin{center}
\normalsize
M. J. E. Richardson\\
{ \it Department of Theoretical Physics, University of Oxford\\ 1
Keble Road, Oxford, OX1 3NP, UK.\\[3mm] {\verb+mjer@thphys.ox.ac.uk+}}
\end{center}
{\bf Abstract:} 
The reaction process $A+B\rightarrow \emptyset$ is modelled for
ballistic reactants on an infinite
line with particle velocities $v_A=c$ and $v_B=-c$ and initially segregated conditions, i.e. all $A$ particles to
the left and all $B$ particles to the right of the origin. Previous
models of ballistic annihilation have particles that always react
on contact, i.e. pair-reaction probability $p=1$. The evolutions of such systems are wholly determined by the
initial distribution of particles and therefore do not have a
stochastic dynamics. However, in this paper the generalisation is made
to $p\leq1$, allowing particles to pass through each other without
necessarily reacting. In this way, the $A$ and $B$ particle domains
overlap to form a fluctuating, finite-sized reaction
zone where the product $\emptyset$ is created. Fluctuations are also included in the currents of $A$ and $B$
particles entering the overlap region, thereby inducing a
stochastic motion of the reaction zone as a whole. These two types of fluctuations,
in the reactions and particle currents, are characterised by the
{\it intrinsic reaction rate}, seen in a single system,
and the {\it extrinsic reaction rate}, seen in an average
over many systems. The intrinsic and extrinsic behaviours are examined
and compared to the case of isotropically diffusing reactants.
\\[5mm]
\date 
\noindent {\bf Journal ref:} {\it J. Stat. Phys.} {\bf 89} (1997) 777 
\\[3mm]
\rule{7cm}{0.2mm}
\begin{flushleft}
\parbox[t]{3.5cm}{\bf Key words:} ballistic annihilation, reaction
process, reaction zone, non-equilibrium statistical mechanics, exactly
solved model
\parbox[t]{12.5cm}{ }
\\[2mm]
\parbox[t]{3.5cm}{\bf PACS numbers:} 02.10.Eb, 05.40.+j, 05.70.Ln, 64.60.Cn
\\[2mm]
\parbox[t]{3.5cm}{\bf Short title:} Exact Solution of Two-Species
Ballistic Annihilation
\end{flushleft}
\normalsize
\thispagestyle{empty}
\mbox{}
\pagestyle{plain}
%
%
%
\newpage
\setcounter{page}{1}
\setcounter{equation}{0}

\nsection{INTRODUCTION}
A detailed understanding of reaction systems is an essential
ingredient for the study of a broad range of problems
\cite{general}. In systems of many interacting reactants, it
often happens that knowledge of the precise physical
mechanism whereby reactions occur is irrelevant in determining the macroscopic
behaviour. The most
important factors are the
number of different reacting species that combine in a single reaction, the type of
motion that each of these species performs and the initial/boundary
conditions in the vessel
containing the reactants. Thus, simple theoretical models that mimic
these elements can be used to study real systems as diverse as exciton dynamics in polymer chains
\cite{ED}, monopole annihilation in the early universe \cite{MM} as
well as the more conventional chemical processes \cite{chem}. However, even such
highly simplified models can prove difficult to analyse. This is
especially true if the reactions are not fully reversible as the
methods of equilibrium statistical mechanics cannot be used, due to the lack of detailed balance.

Many different, idealised reaction systems have been studied
analytically over the last few decades \cite{general} -- \cite{IKR} usually for the case of fully irreversible
reactions. These include same species ($nA\rightarrow \emptyset$) and
multi-species ($\sum_kn_kA_k\rightarrow \emptyset$) processes, where $n_k$ of each
distinct $A_k$ species combine to form a single inert product
$\emptyset$. The multi-species reaction systems exhibit particularly
rich behaviour because reactions can only occur at places where all
the necessary constituents are present - so called {\it reaction
zones}. Often, due to the initial/boundary conditions or spontaneous
symmetry breaking (\cite{BL},\cite{LC}) each species of reactant is largely confined to its
own domain. The places where these domains overlap usually takes up
only a small part of the whole system, often causing the net reaction rate to
differ drastically from that predicted by the mean-field-like rate equation. Such
behaviour can give rise to complicated structures, and is of
particular interest in the context of pattern formation and growth
determination in organisms \cite{turing}. The physics of multi-species
reaction processes is, therefore, largely determined by
activity in and around the reaction zones. The important factors being
the way reactants flow from their respective domains into the overlap region as well as the
behaviour of the reaction mix inside the reaction zone. 

In this paper a model of the two-species system $A+B\rightarrow \emptyset$ is
introduced for the case of ballistically moving reactants and a
pair-reaction probability less than one. The model is then solved exactly,
allowing an analytic study of the dynamic reaction zone formed at the
overlap of the $A$ and $B$ domains. However, before describing the
model in detail two important statistical quantities associated with
the reaction zone are defined.
\newpage

\nsubsecnn{The intrinsic and extrinsic reaction rates}
A comprehensive analysis in the RG framework for the two-species
$A+B\rightarrow \emptyset$ reaction system with {\it diffusive} reactants \cite{BHC} identified two
distinct sources of fluctuations affecting the behaviour of the
reaction zone, first inside the reaction zone itself and second in
the currents of particles entering the reaction zone. 

The shape and size of the reaction zone is determined by the typical
lifetimes and motions of reactants in the overlap region. These lifetimes
in turn depend on the fluctuating local densities of reactants, as
well as the reaction probability for an interacting pair of $A$ and $B$ particles. As the density fluctuations will be caused (in part)
by reactions that have already occurred, the reaction rate can become
highly correlated both spatially and temporally. This behaviour in the overlap region leads to the
following definition for the {\it intrinsic reaction rate}.
\begin{itemize}
\item The {\it intrinsic reaction rate} ${\cal R}_I(x_r,t)$ measured at
a distance $x_r$ from the centre of the overlap region at time $t$, is
the typical spatial reaction rate, i.e. production rate of $\emptyset$
particles per unit time per unit space, seen in a single realisation
of the system's evolution.  The intrinsic reaction rate characterises
the {\it intrinsic reaction zone} - the instantaneous reaction zone
formed where the $A$ and $B$ domains overlap.
\end{itemize}
A second source of fluctuations can come from the $A$ and $B$ currents that
flow into this (intrinsic) reaction zone. Even if these currents are (on average)
equal in magnitude, any fluctuations about the average will contribute a noisy component to the reaction zone's motion, i.e. it will
move stochastically about its expected position. Hence, given
some initial conditions only a probabilistic statement can be made
about the reaction rate at some later place and time $(x,t)$. This suggests the
definition of a second quantity, the {\it extrinsic reaction rate}.
\begin{itemize} 
\item The {\it extrinsic reaction rate} ${\cal R}_E(x,t)$ is defined
as the probability density for reactions to occur at a time $t$
and position $x$ given some initial distribution function for the
particles' positions. Therefore, ${\cal R}_E$ is the expected reaction
rate found by averaging over all allowed realisations of the system's
evolution.
\end{itemize}

Models with ballistically moving reactants have
been extensively studied in the context of the reaction kinetics in an ideal
gas, where the mean free path of the reacting particles is similar to their
separation \cite{EF}--\cite{Pias}. In particular the two-species case,
with particles $A$ and $B$ having velocities $v_A=+c$ and $v_B=-c$ respectively,
has been studied with {\it homogeneous} initial conditions, i.e. each species
initially randomly placed throughout the infinite line. More recently, the
initial conditions of {\it segregated} particles was studied
(\cite{Droz1},\cite{Droz2}) with the $A$s initially to the left and the $B$s to the right of
the origin, again with reactants that always annihilate on contact. The initial
positions for each species of particle (confined to their own domains on
either side of the origin) were chosen to be random, thereby introducing
fluctuations into the particle currents flowing into the reaction zone. By
averaging over all initial distributions the
form of the extrinsic reaction rate was derived and shown to be Gaussian. However,
because there can be no reaction fluctuations in this model (the
particles always annihilate on contact) the $A$ and $B$
domains never overlap and the instantaneous reaction zone has effective width
zero. Also, as the reactants move ballistically the system does
not have a genuinely stochastic evolution, i.e. once the initial conditions
(the particle positions) are fixed so is the system's future. 

This model has a generalisation to the case of arbitrary reaction
probability; this is the general model introduced below. The
fluctuations in the particle streams can be retained and combined with
the fluctuations in the (now) finite-sized overlap region. This then
allows the
intrinsic and extrinsic behaviour of the $A+B\rightarrow \emptyset$
reaction system to be studied analytically.

\nsubsecnn{The definition of the model}
The model consists of a one-dimensional continuous space in which two
species of reactants, $A$
and $B$ particles, move with fixed velocities $v_A=+c$ and
$v_B=-c$, see figure 1. Initially, the reactants are separated,
i.e. at time
$t=0$ $A$ particles are distributed in the interval $[-\infty:0]$ at
positions $(y_1,y_2\cdots)$ and the $B$ particles are in the
interval $[0:\infty]$ at positions $(z_1,z_2\cdots)$. The subscripts on $y_m$ and $z_n$ refer to the
relative {\it initial} order of the particles counted from the origin.  Because of the ballistic motion of the reactants, the trajectories
of the particles retain their initial ordering for all time. 

Two different
distributions for $\{y\}$ and $\{z\}$ will be considered, first equally-spaced reactants (section III) and second, random
positions of the reactants (section IV). For both cases the average
density is chosen to be $\varrho$, leading to average particle currents of $\pm
c\varrho$. However, the initial conditions studied in section IV will be shown
to introduce Gaussian fluctuations about these average values. 

The position $x$ and time $t$ that the $m$th $A$ and $n$th $B$
particles' trajectories intersect are 
\begin{eqnarray}
x=(y_m+z_n)/2 & & t=(z_n-y_m)/2c \label{traji}
\end{eqnarray}
When such an $(m,n)$ pair of reactants' trajectories meet, there are
three distinct events that can occur.
\begin{itemize}
\item If both reactants still occupy their trajectories a
reaction can occur with probability $p$. If a reaction occurs both
particles are removed from the system and a $\emptyset$ product
particle is considered to have been deposited at the point of
annihilation. The $\emptyset$ merely serves as a marker and plays no
further role in the evolution of the system.
\item If both reactants are still travelling along their trajectories then
with probability $q=(1-p)$ no reaction occurs and the particles continue unaffected.
\item If one of the particles has previously been annihilated,
i.e. one trajectory is unoccupied, the other particle continues
unaffected with probability 1. (Of course, if neither particle is
present no change occurs when the empty trajectories cross.)
\end{itemize}
Thus, two distinct sources of fluctuations are included in this model, in the
reactions (if $p<1$) and in the currents of particles flowing
into the reaction zone (if $\{y\}$ and $\{z\}$ have random
elements). Hence, the system models a finite-sized stochastically moving reaction
zone, with measurable intrinsic and extrinsic behaviour. However, it is clear that this model represents a special case, in that the two
sources of noise are uncoupled. The fluctuations in the trajectories are quenched at $t=0$, and the probability that an
$(m,n)$ pair annihilates depends only on the total number of
trajectories crossed by each particle, and not the trajectories' exact
positions. This uncoupling of fluctuations allows the probability
density ${\cal D}$ for an $(m,n)$ pair to
react at $(x,t)$ to be written as the product of two independent distributions
\begin{eqnarray}
{\cal D}(m,n,x,t)&=&P(m,n){\cal G}_{mn}(x,t) \nonumber
\end{eqnarray}
where $P(m,n)$ is the probability for an $(m,n)$ pair to mutually
annihilate and ${\cal G}_{mn}(x,t)$ is the probability density for the
trajectory intersections (\ref{traji}). The calculation of the total annihilation
probability can therefore be decomposed into, (i) a counting problem
for the integer variables $(m,n)$, and (ii) the derivation of the distribution functions for the continuous random numbers $\{y\}$ and
$\{z\}$.

The remainder of the paper is structured as follows. In section II the
probability of pair annihilation $P(m,n)$ is derived (\ref{P2}) by mapping the model onto a simple system of target
particles on a one-dimensional lattice. The behaviour of the lattice
system is then briefly examined in the context of radiation damage of
crystals. In section III the form of the (intrinsic) reaction rate for the simple
case of equally-spaced reactants is studied. In particular, the
steady-state reaction-rate (\ref{intrinsz}) and particle densities (\ref{profz}), and
the time-dependence of the reaction rate (\ref{Papp2}) are
derived. Finally, in section IV fluctuations in the initial particle
positions are treated. The distribution function ${\cal G}_{mn}(x,t)$ is 
found and used to derive the forms of the intrinsic and extrinsic
reaction rates, (\ref{INT}) and (\ref{EXT}) respectively. The appendix
shows how the calculation in section II may be translated into the
second-quantisation formalism and relates the mapped system to the
algebra $SU_q(2)$.

\nsection{THE PAIR-REACTION PROBABILITY}
The main result of this section is the derivation of the $(m,n)$ pair
annihilation probability $P(m,n)$, equation (\ref{P2}). In section I, it
was noted that this quantity is independent of the initial positions
of the particles, depending only on their relative order. Making use
of this fact, a simpler system can be treated, that still preserves the
order that the $A$ and $B$ particles pass through each other.

Consider now a one-dimensional, semi-infinite lattice (with sites
$n=1,2\cdots$) with a $B$ particle initially occupying each site. At discrete time steps ($m=1,2\cdots$) the $m$th $A$ particle is
`shot' through the $B$ array, passing sequentially through each
site until its eventual annihilation, see figure 1. An $A$ particle either moves through a site with a
$B$ present with probability $q$, or reacts with a $B$ at that site
with probability $p$.  Once an $A$ and $B$ pair have reacted, the site that the $B$
occupied becomes vacant and any subsequent $A$ that passes
through the vacant site does so with probability $1$. This whole
process, of a single $A$ passing through the lattice and eventually
reacting, is considered to happen in a negligible amount of time. This
mapped system is merely a deformation of the coordinates used in the
original model, so both systems share the same pair
annihilation probability $P(m,n)$. 

It is
useful to consider the statistics of the positions of the $m$
vacancies that exist in the lattice immediately after the $m$th time step.  The
positions of these vacancies are labelled $(n_1\cdots n_m)$, where it
is important to note that the subscript used refers to the relative
positions of the sites, i.e. $n_1<n_2<\cdots<n_m$, and not which $A$
particle caused the vacancy at that position.

The rest of this section is devoted to the calculation of $P(m,n)$,
the probability that the $m$th $A$ reacts at site $n$. First, the
probability of a particular distribution $\Psi_m(n_1\cdots n_m)$ for
the positions of the $m$ vacancies is derived, equation (\ref{psiM}). These probabilities can be used as a convenient basis, in the sense that all other
probabilistic quantities may be expressed as linear combinations of
the $\{\Psi_m\}$. This basis is then used to find the expected vacancy density
$V_m(n)$ at site $n$, after $m$ $A$ particles have passed,
equation(\ref{prof}). As an aside, a parallel is drawn between this
simplified model and a crystal that has been damaged by radiation. In
particular, it is shown that the damaged region described by $V_m(n)$
propagates like a soliton through the $B$ array, equation
(\ref{shock1}). Finally, the discrete gradient of $V_m(n)$ is then used to calculate the required quantity $P(m,n)$, equation
(\ref{P2}). The method
described below translates into the second-quantisation formalism and
shows the system to be described by the algebra $SU_q(2)$. The
techniques used in this formalism are briefly reviewed in the appendix.

\nsubsecnn{The basis for the $m$th time step $\Psi_m$}
The probability $\Psi_m(n_1\cdots n_m)$ that after $m$ $A$ particles have
passed through the $B$ lattice vacancies exist at sites
$n_1\!\!<\!\!n_2\!\!<\!\!\cdots\!\!<\!n_m$ is now derived. Consider
first the simplest case $m=1$, i.e. just after the first time
step. The probability that the $A$
particle has annihilated with a $B$ on site $n_1$ producing a vacancy
there is
\begin{eqnarray}
\Psi_1(n_1)&=&pq^{{n_1}-1}=(q^{-1}-1)q^{n_1} \nonumber
\end{eqnarray}
where $p$ is the probability a single reaction could occur, and $q=(1-p)$. 

Now consider the state of the system after the second time step,
i.e. after a total of two $A$ particles have passed through the $B$
lattice. Vacancies now exist at sites $n_1$ and $n_2$ (where
relabelling may be necessary to ensure that $n_1<n_2$). There are two
histories that contribute to this configuration, each with
different probabilities. Either a vacancy first appeared at site $n_2$
and then the second one at site $n_1$, or the first vacancy at site $n_1$ and the second at site $n_2$. The probability
of the first history is just the product $\Psi_1(n_1)\Psi_1(n_2)$ as
the second $A$ particle does not pass through the vacancy
produced by the first $A$ particle. However, in the
second history described, the second $A$ particle does pass through
the vacancy at $n_1$, increasing its chance of reacting with
any site $n>n_1$ by a factor of $q^{-1}$. Therefore,
\begin{eqnarray}
\Psi_2(n_1,n_2)&=&\Psi_1(n_1)\Psi_1(n_2)+q^{-1}\Psi_1(n_1)\Psi_1(n_2) \nonumber\\
\Psi_2(n_1,n_2)&=& (1+q^{-1})\Psi_1(n_1)\Psi_1(n_2) \label{psi2}\\
\Psi_2(n_1,n_2)&=&(q^{-2}-1)(q^{-1}-1)q^{n_1+n_2}. \nonumber
\end{eqnarray}
It is simple to generalise to the $m$-vacancy basis $\Psi_m(n_1\cdots n_m)$. The states at the $(m-1)$th time step that
can contribute to an $m$th time step configuration are those with
vacancies at $(n_2\cdots n_m)$, $(n_1,n_3\cdots n_m)$, $\cdots$, $(n_1\cdots n_{m-1})$. In
each of these cases, to produce the final state $(n_1\cdots n_m)$ the
$m$th $A$ particle must react with
sites $n_1$,$n_2\cdots n_m$ respectively. Therefore, taking account of
how many vacancies the $m$th $A$ particle must pass through in each
case, a relation between the bases that describe the system after time steps $(m-1)$ and $m$ can be written
\begin{eqnarray}
\Psi_m(n_1\cdots n_m) &=& \Psi_1(n_1)\Psi_{m-1}(n_2\cdots n_m)\nonumber\\
&&+q^{-1}\Psi_1(n_2)\Psi_{m-1}(n_1,n_3\cdots n_m)+\cdots \nonumber\\
&&+q^{-(m-1)}\Psi_1(n_m)\Psi_{m-1}(n_1,n_2\cdots n_{m-1}).\nonumber
\end{eqnarray}
Assuming that it is possible to write
$\Psi_{m-1}=C_{m-1}\prod_{j=1}^{m-1}\Psi_1(n_j)$ and using the
result (\ref{psi2}), the form for general $m$ follows by induction
\begin{eqnarray}
\Psi_m(n_1\cdots n_m) &=&
(1+q^{-1}+\cdots q^{-(m-1)})C_{m-1}\prod_{j=1}^m\Psi(n_j)=C_m\prod_{j=1}^m\Psi(n_j)\nonumber
\\
\Psi_m(n_1\cdots n_m)&=&\prod_{j=1}^{m}\left[\frac{q^{-j}-1}{q^{-1}-1}\Psi_1(n_j)\right]=\prod_{j=1}^{m}\left[(q^{-j}-1)q^{n_j}\right].
\label{psiM}
\end{eqnarray}
Hence, the distribution of the $m$ vacancies is given by a product of
exponentials in the site labels, with care being taken to preserve the order
$n_1\!\!<\!\!n_2\!\!<\!\!\cdots\!\!<\!n_m$ in any sums that they
appear.
 
\nsubsecnn{The vacancy density $V_m(n)$}
The probability that a vacancy is found at site $n$ after $m$ $A$
particles have passed through the lattice, $V_m(n)$, is now
derived. This density is given by the sum of all the $\Psi_m$
that include a vacancy at the site $n$,
\begin{eqnarray}
V_m(n)&=&\sum_{k=1}^{m}\tilde{\Psi}_m(n_1\cdots,n_k=n,\cdots n_m) \nonumber
\end{eqnarray}
where the notation $\tilde{\Psi}$ is used to denote an internal sum
over all the unfixed variables, i.e. any $n_j$ with $j\not=k$,
that respects the order $n_1\!\!<\!\!n_2\!\!<\!\!\cdots\!\!<\!n_m$. Hence,
\begin{eqnarray}
\tilde{\Psi}_m(n_1\cdots,n_k=n,\cdots n_m)=&& \nonumber \\
\sum_{n_1=1}^{n-(k-1)}\cdots\sum_{n_{k-1}=n_{k-2}+1}^{n-1}&&\sum_{n_{k+1}=n+1}^{\infty}\cdots\sum_{n_{m}=n_{m-1}+1}^{\infty}\Psi_m(n_1\cdots,n_k=n,\cdots
n_m). \label{sums}
\end{eqnarray}
For the case $m=1$, the vacancy density is simply $V_1(n)=\Psi_1(n)$. However, after the second $A$ particle passes
through the $B$ array a sum must be made over the unfixed variables
\begin{eqnarray}
V_2(n)&=&\tilde{\Psi}_2(n_1=n,n_2)+\tilde{\Psi}_2(n_1,n_2=n) \nonumber \\
V_2(n)&=&\sum_{n_2=n+1}^{\infty}\Psi_2(n_1=n,n_2)+\sum_{n_1=1}^{n-1}\Psi_2(n_1,n_2=n)
\nonumber\\
V_2(n)&=&(1+q^{-1})\Psi_1(n)\left(\sum_{j=n+1}^\infty\Psi_1(j)+\sum_{j=1}^{n-1}\Psi_1(j)\right)\nonumber\\
V_2(n)&=&q^n(q^{-2}-1)\left(\sum_{j=1}^{\infty}\Psi_1(j)-\Psi_1(n)\right)\nonumber\\
V_2(n)&=&q^n(q^{-2}-1)\left(1-V_1(n)\right). \nonumber
\end{eqnarray}
The density for $m=2$ is therefore related to the $m=1$ case, by making
use of the product form of $\Psi_2$. It is
possible to generalise this result and obtain a recursion
relation. The following two results will prove useful. First the
product form of $\Psi_m$ is used to relate $\Psi_m$ to $\Psi_{m-1}$
\begin{eqnarray}
\Psi_m(n_1\cdots,n_k=n,\cdots
n_m)&\equiv&q^n(q^{-m}-1)\Psi_{m-1}(n_1\cdots n_{k-1},n_{k+1}\cdots
n_m). \label{result1}
\end{eqnarray}
Second, a slightly less trivial result, that nevertheless has a simple interpretation
\begin{eqnarray}
\sum_{k=1}^{m}\tilde{\Psi}_{m-1}(n_1\cdots n_{k-1},n_{k+1}\cdots
n_m)&\equiv&1-\sum_{k=1}^{m-1}\tilde{\Psi}_{m-1}(n_1\cdots,n_k=n,\cdots
n_{m-1}) \label{result2}\\
&\equiv&1-V_{m-1}(n). \nonumber
\end{eqnarray}
The LHS of this equation is the sum of all the $\Psi_{m-1}$ that do
not include a vacancy on site $n$. This is simply the sum of all
possible $\Psi_{m-1}$ ($\equiv1$ by normalisation) less those that
include the site $n$; the statement on the RHS of
(\ref{result2}). The $\tilde{\Psi}$s in this equation are identical to
those in equation (\ref{sums}) in as much as they involve an ordered
sum over all unfixed variables. However, it should be noted that,
though on the LHS the term $n_k=n$ has been factored out by using
(\ref{result1}), the order restriction still holds, i.e. $n_{k-1}<n$
and  $n_{k+1}>n$. 

Both these results generalise the method used
already to find $V_2(n)$. Therefore, for the case of $m$ vacancies
\begin{eqnarray}
V_m(n)&=&\sum_{k=1}^{m}\tilde{\Psi}_m(n_1\cdots,n_k=n,\cdots n_m)\nonumber\\
V_m(n)&=&q^n(q^{-m}-1)\sum_{k=1}^{m}\tilde{\Psi}_{m-1}(n_1\cdots
n_{k-1},n_{k+1}\cdots n_m) \label{step2}\\
V_m(n)&=&q^n(q^{-m}-1)\left(1-V_{m-1}(n)\right). \label{rhorec}
\end{eqnarray}
The recursion relation (\ref{rhorec}) can be
solved, with the initial condition $V_1(n)=\Psi_1(n)$, to give the
vacancy density at site $n$ after $m$ $A$ particles have
passed through the $B$ array
\begin{eqnarray}
V_{m}(n) &=&
-\sum_{j=1}^{m}\left[\prod_{k=j}^{m}(q^n-q^{n-k})\right]
\label{prof} \\
V_{m}(n) &=&
-q^{(n-m)}\sum_{j=0}^{m-1}\left[q^{(n-m)j}\prod_{k=0}^{j}(q^m-q^k)\right].
\nonumber
\end{eqnarray}
Interestingly, in the limit of large $m$ (in particular $q^m\ll p$) the vacancy
profile depends purely on the difference $s=(n-m)$. Interpreting the
system as a crystal (the $B$ array) that is being damaged by incoming
radiation (the $A$ particles), this implies that
the interface between the damaged and undamaged regions reaches a
steady-state moving profile
\begin{eqnarray}
\lim_{m\rightarrow\infty}V_m(n)=V(s)&=&q^s\sum_{j=0}^{\infty}(-q^s)^jq^{j(j+1)/2}
\label{shock1} \\
&\sim& \frac{1}{2}\left(1-\tanh(ps/2)\right)+O(p) \label{shock2} 
\end{eqnarray}
where the second equation becomes valid in the case of low reaction
probability $p\ll1$ and is analytically continued for $s<0$. Therefore, the damaged region propagates like a soliton
through the $B$ array.

\nsubsecnn{The pair-reaction probability $P(m,n)$}
The difference between the vacancy density at site $n$ 
just after the
$m$th and $(m-1)$th time steps is the probability that at the $m$th
time step an annihilation occurs at site $n$. This is the required pair
reaction probability $P(m,n)$.
\begin{eqnarray}
P(m,n) &=& V_{m}(n)-V_{m-1}(n) \nonumber\\ 
P(m,n) &=&
q^{(n-m)}\sum_{j=0}^{m-1}\left[q^{(n-m)j}(1-q^{j+1})\prod_{k=1}^{j}(q^m-q^k)\right],
\label{P2}
\end{eqnarray}
which is the main and final result of this section. For the case
studied in \cite{Droz1}, i.e. $q=0$, this result reduces to a delta function as
expected, i.e. $P(m,n)=\delta(m,n)$. In the next section $P(m,n)$ is used
to study the original model described in section I, with equal
spacing between neighbouring reactants at $t=0$.

\nsection{THE REACTION ZONE WITHOUT CURRENT FLUCTUATIONS} 
Reverting back to the original model described in the first section, $P(m,n)$ is
reinterpreted as the probability that an $(m,n)$ pair of particles
mutually annihilates in
a system of ballistically moving reactants. In this section, the
model is studied with the
initial conditions $y_m=-m\varrho^{-1}$ and
$z_n=n\varrho^{-1}$, i.e. with no fluctuations in the particle
currents. The point of intersection of an $(m,n)$ pair of trajectories is therefore
\begin{eqnarray}
x=x_r=(n-m)/2\varrho& &t=(m+n)/2c\varrho, \label{easy}
\end{eqnarray}
i.e. reactions occur only at discrete positions and times. The lack of fluctuations in the
currents negates the need to discuss the extrinsic behaviour, as the
intrinsic reaction zone will not wander stochastically. In fact as ${\cal
R}_I={\cal R}_E$ for these initial conditions, the zero current-fluctuation reaction zone will be denoted
by ${\cal R}_0$ to avoid confusion with section IV. The steady currents also mean that the centre of the overlap region is always at the origin,
and therefore $x_r=x$ in this case. As will be shown in the next
section, even when current fluctuations are taken into
account, the form of the intrinsic reaction rate remains the same as for
this case, as long as $x_r$ is then measured from the
centre of the stochastically moving overlap centre seen in section IV. 

In the rest of this section several aspects of equation (\ref{P2})
are studied. First, the steady-state reaction rate
(\ref{intrinsz}) and particle densities are derived (\ref{profz}). Second, the dynamics of $P(m,n)$ in the $O(p)$
approximation is studied and the system is shown to decay to the
steady state in finite time (\ref{Papp2}). Finally,
the correlations in the reaction zone are briefly discussed, with reference to the
mean-field and $O(p)$ approximations.

\nsubsecnn{The steady-state limit and $O(p)$ time dependence}
After a sufficient length of time, characterised by $q^m\ll p$,
equation (\ref{P2}) becomes a function of $(n-m)=s$ only. This is in much the same way as for
equation (\ref{shock1}) and can also be interpreted as a late-time or steady-state limit. Therefore, for large $m$ (or $n$
by symmetry) any $(m+l,n+l)$ pair will have the same probability of
reacting, regardless of the value $l$ takes,
\begin{eqnarray}
\lim_{m\rightarrow\infty}{P(m,n)}=P(s)&=&q^s\sum_{j=0}^{\infty}\left[(-q^s)^jq^{j(j+1)/2}(1-q^{j+1})\right]
\label{steadyP2} \\
P(s)&=&\frac{p}{4\cosh^2(ps/2)}+O(p^2).   \label {steadyP}
\end{eqnarray}
The production rate of $\emptyset$
particles at a position $x_r=s/(2\varrho)$ can now be found, namely $2c\varrho
P(2x_r\varrho)$. However, the nature of the initial conditions means
that reactions can occur only at discrete positions ($s=0,\pm1\cdots$). So, to derive the
reaction rate ${\cal R}_0(x_r)$ which is the production rate {\it per unit
space}, a coarse-graining is made over a length scale $\varrho^{-1}$
\begin{eqnarray}
{\cal R}_0(x_r)&=&2c\varrho^2P(2x_r\varrho). \label{intrinsz}
\end{eqnarray}
This reaction rate (characterising the intrinsic reaction zone) is shown graphically in figure 2 for $p=q=1/2$ with
comparison made between the $O(p)$ and mean-field approximations (to
be described below). The particle densities can also be found, given that $[1-V(s)]$ is the probability a $B$ particle still exists
just after a reaction occurs at site $s$ (and similarly for the $A$
particles). Therefore, the density profiles $\varrho_A(x_r)$ and $\varrho_B(x_r)$
in the steady-state are
\begin{eqnarray}
\varrho_A(x_r)=\varrho[1-V(-2\varrho x_r)] &\mbox{ and }& \varrho_B(x_r)=\varrho[1-V(2\varrho
x_r)] \label{profz}.
\end{eqnarray}
These profiles are also shown, for the same parameters $q=p=1/2$,
in figure 2, with comparison made to the $O(p)$ approximation. 

To examine the passage to the steady state, it is illuminating to
derive the time dependence of (\ref{P2}) to first order in $p$, i.e. take the limit $p\ll 1$ in equation
(\ref{P2}) but still keep $m$ finite. 
\begin{eqnarray}
P(m,n) &=&p\left(2\cosh(p(n-m)/2)-\exp^{-p(m+n)/2}\right)^{-2} +O(p^2)
\label{Papp} \\
{\cal
R}_0(x_r,t)&=&\frac{pc\varrho^2\theta(c\varrho t-1-\varrho |x_r|)}{2\left(\cosh(p\varrho
x_r)-\exp^{-pc\varrho  t}\right)^2} +O(p^2), \label{Papp2}
\end{eqnarray}
where $\theta(u)=1$ or $0$ for $u\geq0$ or $u<0$, respectively. It
is clear that the system reaches a steady state
exponentially quickly (a characteristic of systems with short-range correlations) with a decay time $(pc\varrho)^{-1}$. To leading order
in $t$, this exponential decay time will also be a feature of the model
with randomly
distributed particles to be examined in section IV. 

\nsubsecnn{Correlations, the $O(p)$ and mean-field approximations}
The first order in $p$ approximations (\ref{shock2}), (\ref{steadyP}),
(\ref{Papp}) and (\ref{Papp2}) valid for $p\ll 1$
represent the case where each particle passes through very many of its opposite kind
before finally annihilating. In this $p\ll 1$ limit it could therefore be expected that a
{\it mean-field} approach becomes exact. For the reaction model the mean-field
approximation is one that neglects density
correlations in the reaction zone. It assumes that the probability of
a reaction occurring depends on the product of the probabilities that
an $A$ and $B$ particle are present, i.e. that the $A$ and $B$
densities are independent statistical quantities. Therefore, given that $[1-V_{m-1}(n)]$ is the probability that the $n$th $B$ has
not annihilated before meeting the $m$th $A$ particle (and similarly
for the $m$th $A$ particle by symmetry), the mean-field approximation is
\begin{eqnarray}
P_{mf}(m,n)&=&p[1-V_{n-1}(m)][1-V_{m-1}(n)] \nonumber\\
P_{mf}(s)&=&p[1-V(s-1)][1-V(s+1)] \label{steadymf}
\end{eqnarray}
where the second equation is the steady-state limit. From these
equations it can be seen by substitution that the
$O(p)$ approximations (valid only in the limit $p\rightarrow 0$) are indeed mean-field like. Though, this is not
to say that the $O(p)$ and mean-field approximations are equivalent
outside the small $p$ limit, as can been seen in figure 2 where
comparison can be made for $p=0.5$. 

For the $p\ll1$ case, the reaction zone will be
much larger than the interparticle spacing, and any structure will slowly vary
with the variables $m$ or $n$. Therefore, the $O(p)$ approximations can also be captured from
the {\it continuum} mean-field equations
\begin{eqnarray}
\frac{\partial \varrho_A}{\partial t}=-c\frac{\partial
\varrho_A}{\partial x}-r\varrho_A\varrho_B & & \frac{\partial \varrho_B}{\partial
t}=+c\frac{\partial \varrho_B}{\partial x}-r\varrho_A\varrho_B. \nonumber
\end{eqnarray}
where $r$ is the reaction parameter. However, the full results (\ref{prof}), (\ref{shock1}), (\ref{P2}) and
(\ref{steadyP2}) cannot be obtained by such a mean-field approach, as
can be seen in figure 2 where $P_{mf}(s)$ and $P(s)$ are compared. The
figure shows that the correlations are strongest in the centre of the
reaction zone and decay quickly towards the edge of the overlap region.

Interestingly, a similar cross-over to mean-field behaviour is also seen in the case of
diffusive reactants \cite{BHC} in the limit of high diffusivity and
low reaction rate. The interpretation, of many interactions between
particles before
final annihilation occurs, is the same.

\nsection{REACTION ZONES}
In this section the case of fluctuations in the initial conditions is treated (though still
with each species initially segregated to either side of the origin). This allows the
intrinsic and extrinsic reaction rates for the system with fluctuations in both reactions and particle
currents to be found, equations (\ref{INT}) and (\ref{EXT}). 

The initial positions of each species of particle are chosen to be
uncorrelated with an average spacing $\varrho^{-1}$ and
are described by Poissonian distributions with the variances
of the interparticle spacing being $\varrho^{-2}$. Hence, both $y_m$ and $z_n$, the initial positions of the $m$th $A$
and $n$th $B$ particles, can be written as sums of
independent random numbers
\begin{eqnarray}
y_m &=&
(y_{m}-y_{m-1})+(y_{m-1}-y_{m-2})+\cdots+(y_{2}-y_{1})+(y_{1}-0) \nonumber\\ 
z_n &=& (z_{n}-z_{n-1})+(z_{n-1}-z_{n-2})+\cdots+(z_{2}-z_{1})+(z_{1}-0).
\nonumber
\end{eqnarray} 
The central limit theorem can now be used, stating that for
large $m$ and $n$ such sums of many independent random numbers, drawn from the same
distributions, become the Gaussians
\begin{eqnarray}
{\cal Y}(y_m)\propto\exp{\left(-\frac{(\varrho y_m+m)^2}{2m}\right)} &
& {\cal Z}(z_n)\propto\exp{\left(-\frac{(\varrho z_n-n)^2}{2n}\right)}. \label{GAUSSES}
\end{eqnarray}
The probability density for the $m$th $A$ and $n$th $B$
trajectories to intersect at position $x$ and time $t$ (the function
${\cal G}_{mn}(x,t)$ defined in section I) is now
easily calculated. As both $x$ and $2ct$ are linear combinations
of independent Gaussian variables, for $s\ll m$ their distribution functions are
also Gaussians with means and variances linear combinations of those in
equation (\ref{GAUSSES}) 
\begin{eqnarray}
{\cal G}_{mn}(x,t)=\frac{2c\varrho^2}{\pi(n+m)}\exp\left(-\left[\frac{(2\varrho
x-(n-m))^2}{2(m+n)}+\frac{(2c\varrho t-(m+n))^2}{2(m+n)}\right]\right).  \label{GAUSSES2}
\end{eqnarray}

Before calculating the intrinsic and extrinsic reaction rates for
$p<1$, the deterministic case $p=1$ examined in \cite{Droz1} is first
reviewed. Because in this case, the reaction probability
$P(m,n)=\delta_{mn}$ the two domains of $A$ and
$B$ particles never overlap and only an extrinsic reaction rate
can be meaningfully defined. The probability density for
reactions to occur, i.e. for $(m,m)$ pairs to intersect, at $(x,t)$ is given by
\begin{eqnarray}
{\cal
R}_{p=1}&\simeq&\sum_{m=1}^\infty\sum_{n=1}^{\infty}\delta_{mn}{\cal
G}_{(m,n)}(x) \nonumber\\
{\cal R}_{p=1}&\simeq&\frac{c\varrho^2}{\sqrt{\pi
c\varrho t}}\exp\left(-\frac{(\varrho x)^2}{c\varrho t}\right) \label {droz}
\end{eqnarray}
in the limit of large $m$. This is the extrinsic reaction rate for the $p=1$ model and it implies the reaction front is a Gaussian random
walker, covering a typical distance $\sim(ct/\varrho)^{1/2}$ in a
time $t$. 

For the general case $p\leq1$ the points of intersection of these $(m,m)$ pairs will be
used as a convenient definition for the centre of the stochastically
moving reaction zone. The forms of the intrinsic and extrinsic reaction rates can now be calculated. 

\nsubsecnn{The intrinsic reaction rate}
The intrinsic reaction rate as defined in section I, is the reaction-rate profile seen
in a single realisation of the system's evolution (measured relative to the centre of
the $A$ and $B$ domains' overlap region). In this section it will be
argued that, if reactions are measured relative to the position of the
$(m,m)$ pair intersections, the intrinsic reaction rate is on average
equal to ${\cal R}_0$, the reaction rate for the zero current-fluctuations case examined in section III.

Consider the set of intersecting particle pairs that can be written
$(m-k,m+k)$ with $m$ fixed and $k$ varying. These pairs all share the same
average time of intersection. Therefore, it is convenient to define
the relative coordinates of intersection $(x_r,t_r)$ of these pairs to the central
$(m,m)$ pair
\begin{eqnarray}
2x_r=(y_{m-k}+z_{m+k})-(y_m+z_m) & & 2ct_r=(z_{m+k}-y_{m-k})-(z_m-y_m) \nonumber\\
x_r=k\varrho^{-1}+O(k^{1/2}\varrho^{-1}) & & t_r=0+O(k^{1/2}(2c\varrho)^{-1}) \label{ref}\nonumber
\end{eqnarray}
where the $O(k^{1/2})$ deviations are those expected from the Gaussian
fluctuations (\ref{GAUSSES}). As the deviations are not
functions of $m$, the noise in the particles' initial positions does not
produce any time-dependent dispersion, i.e. the statistics of the pair
intersections in the overlap region reach a steady-state. The relative time and positions
are on average the same for the equally-spaced case seen
in section III, though the Gaussian fluctuations about these average values
introduce a $\sim k^{1/2}\varrho^{-1}$ uncertainty in the position of the
reactions. However, this broadening is not
sufficiently strong to disrupt the ${\cal R}_0$ profile if the
inequality $p^{1/2}\ll1$ is satisfied, i.e. if the width of the reaction
zone is much greater than the broadening. Rather, the fluctuations act to
smooth the discontinuous nature of the equally-spaced reactant
currents, for which a coarse-graining was necessary in section III. Therefore, from equation (\ref{Papp2}) the steady-state intrinsic reaction rate is
\begin{eqnarray}
{\cal{R}}_I(x_r)&=&2c\varrho^2P(2\varrho x_r) \label{INT} \\
{\cal{R}}_I(x_r)&=&\frac{pc\varrho^2}{2\cosh^2(p\varrho x_r)}+O(p^2) \nonumber
\end{eqnarray}
It also follows that the particle streams have the coarse-grained
density distribution seen in the ordered case. Hence, figure 2 also
represents the profiles seen in the case with Gaussian fluctuations in
the particle currents. 

The width of the intrinsic reaction zone is
$\sim(p\varrho)^{-1}$ and from (\ref{droz}) in a time $\Delta T$ it will typically move
(as a whole) a distance $\sim(c\Delta T/\varrho)^{1/2}$. Hence, any
measurement made of the $\emptyset$ production rate must have
$\Delta T\ll(p^2c\varrho)^{-1}$. Such a restriction can
indeed be satisfied, and the $\emptyset$ production rate that (\ref{INT}) predicts
can be clearly seen in a Monte Carlo simulation of a single system,
i.e. the system is self-averaging.

In the above analysis the steady state was assumed for the intrinsic
reaction rate calculation. However, the
arguments also follow through if the time-dependent intrinsic rate
(\ref{Papp2}) is used, for $p\ll1$. The width of the intrinsic reaction zone $W_I$
therefore varies as 
\begin{eqnarray} 
W_I&\propto&(ct) \mbox{ for $t\ll t_I$ } \nonumber \\
W_I&\propto&(p\varrho)^{-1} \mbox{ for $t\gg t_I$} \nonumber
\end{eqnarray} 
where $t_I=(pc\varrho)^{-1}$ is the relaxation time for the intrinsic
rate given in the previous section. Hence, at early times the width increases linearly, until
finally saturating at a finite time-independent width.

It is unlikely that the present model studied is in the same
universality class as the case of diffusive reactants. However, it is interesting to note that the asymptotics of the intrinsic
reaction rates derived for the diffusive case and in the
present case of ballistic reactants are both of
the same form $\sim e^{-a|x|}$, in the steady state. 

\nsubsecnn{The extrinsic reaction rate}
The extrinsic reaction rate, defined as the reaction rate
at $(x,t)$ averaged over all possible evolutions, implies an average
over all the appropriately weighted initial particle positions $\{y\}$ and $\{z\}$. It is
therefore equivalent to the sum of probability densities for any
$(m,n)$ pair to annihilate at
$(x,t)$. After the intrinsic reaction rate has relaxed to its
steady-state limit, i.e. for times $t\gg t_I$, the extrinsic reaction
rate can be expressed
purely as a function of $s=(n-m)$.
\begin{eqnarray}
{\cal R}_E(x,t) &=& \sum_{m}\sum_{n}P(m,n){\cal G}_{mn}(x,t) \nonumber\\
\lim_{t\rightarrow\infty}{{\cal R}_E(x,t)} &=& \sum_{s}P(s){\cal G}_s(x,t). \nonumber
\end{eqnarray}
The function ${\cal G}_s(x,t)$ is the probability density that an
$(m,m+s)$ pair meet at $(x,t)$ for any value of $m$
with $s$ fixed, in the limit $t\gg(c\varrho)^{-1}$
\begin{eqnarray}
{\cal
G}_s(x,t)&=&\lim_{t\rightarrow\infty}{\sum_{m=1}^{\infty}\frac{c\varrho^2}{\pi
m}\exp\left(\frac{-1}{4m}\left((2\varrho x-s)^2+(2c\varrho t-2m)^2\right)\right)}
\nonumber \\
{\cal G}_s(x,t)&=&\frac{c\varrho^2}{\sqrt{\pi
c\varrho t}}\exp\left(\frac{-(2\varrho x-s)^2}{4c\varrho t}\right) \nonumber.
\end{eqnarray}
Therefore, the extrinsic reaction rate can be written as a convolution
of the intrinsic rate over all the allowed paths the intrinsic
reaction zone can take
(each being weighted by a Gaussian).
\begin{eqnarray}
{\cal R}_E(x,t)&=&\frac{c\varrho^2}{\sqrt{\pi
c\varrho
t}}\sum_{s=-\infty}^{\infty}P(s)\exp\left(\frac{-(2\varrho x-s)^2}{4c\varrho
t}\right)
\label{EXT} \\
{\cal R}_E(x,t)&=&{\sqrt{\frac{p^2c\varrho^3}{16\pi
t}}}\sum_{s=-\infty}^{\infty}\frac{\exp\left(\frac{-(2\varrho x-s)^2}{4c\varrho
t}\right)}{\cosh^2{(ps/2)}}+O(p^2).
\nonumber
\end{eqnarray}
A second characteristic time $t_E=(p^2c\varrho)^{-1}$ is now introduced
into the system by the Gaussian noise. This is the timescale on which
the uncertainty in the position of the intrinsic reaction zone
becomes equal to its steady-state width, i.e. the standard deviation
of the position of the reaction zone's centre is of the order $(p\varrho)^{-1}$. For $t\ll t_E$
the extrinsic and intrinsic rates are effectively the same, as there is little broadening. However, for times $t\gg t_E$ the noise in the
particle currents
dominates, the extrinsic reaction rate becomes a Gaussian and the
result \cite{Droz1} is recovered. This is not surprising as for these
times the structure of the
intrinsic zone is unimportant and can be considered a
point-like random walker. The width $W_E$ of the extrinsic reaction
rate as a function of time is therefore
\begin{eqnarray}
W_E&\propto&(ct) \mbox{ for $t\ll t_I$ }\nonumber \\
W_E&\propto&(p\varrho)^{-1}
\mbox{ for $t_I\ll t\ll t_E$}\nonumber \\
W_E&\propto&(ct/\varrho)^{1/2} \mbox{ for $t\gg t_E$} \nonumber
\end{eqnarray}
where $t_I=(pc\varrho)^{-1}$ and $t_E=(p^2c\varrho)^{-1}$. The
asymptotic behaviour (large $x$ values) of the extrinsic reaction rate
also has distinct forms as a function
of time 
\begin{eqnarray}
{\cal R}_E&\sim&\theta(ct-\varrho|x|) \mbox{ for $t\ll t_I$}
\nonumber \\
{\cal R}_E&\sim&e^{-a|x|} \mbox{ for $t_I\ll t\ll t_E$ }
\nonumber \\
{\cal R}_E&\sim&e^{-bx^2/t} \mbox{ for $t\gg t_E$}.\nonumber
\end{eqnarray}
Hence, there is a cross over from exponential to Gaussian behaviour at
late times.

\nsection{DISCUSSION}
In the present work, a new model for the reaction system
$A+B\rightarrow \emptyset$ with ballistic reactants has
been introduced and solved exactly. The model includes two types of noise,
in the reactions (due to the reaction probability being less
than one) and in the currents of particles (due to disorder in the
particles' initial positions). These fluctuations allow some of the characteristics of real
systems to be exhibited, including a fluctuating $A$ and $B$ overlap
region where the reactions occur, and a reaction zone that moves stochastically throughout
the system. Comparison was made between
the present model of ballistically moving reactants and that of
isotropically diffusing reactants \cite{BHC} studied under the RG framework. In particular, it was noted that
the intrinsic reaction rates (the dynamic
reaction region formed between the fluctuating $A$ and $B$ domains)
have the same asymptotic form, though it is unclear if the two systems share
the same universality class. An important physical difference between
these two cases is that, in the ballistic model, particle order
is preserved and interactions between a given pair of particles can occur only
once. It would therefore be interesting to study the case of
reactants that perform biased diffusion, because such a system would interpolate between the cases of
isotropic diffusion and ballistic motion of reactants. As shown in the
appendix, the method outlined in section II
can be translated into the second-quantisation formalism with the vacancy
dynamics described by the $SU_q(2)$ algebra. Using the tools available in this formalism it may be possible to introduce
interaction terms in the evolution operator that break the order of
the particles, thereby introducing a bias-diffusive
component into the otherwise deterministic motion. 

Another case for further study would be to
examine more closely the relation between the present model and the
partially-asymmetric exclusion process (PASEP) with reflecting
boundaries - also described by $SU_q(2)$ \cite{SaSch}. The present model displays short-range correlations, as does the PASEP with
reflecting boundaries. However, the PASEP with open boundaries is known to
have three phases \cite{SANDOW}--\cite{SHDOM}, one of which has
long-range power-law behaviour. It would therefore be
interesting to see if the open system also translates into a
system of reacting particles.

\nsubsecnn{Acknowledgements}
The author thanks S. Cornell for outlining the system studied in this
paper. J. Cardy, M. R. Evans, R. B. Stinchcombe, G. M Sch\"utz and
P. Frantsuzov are also thanked for useful discussions. Financial
support is acknowledged from the EPSRC under Award No. 94304282.
\newpage
\noindent{\bf APPENDIX. ``Second-quantisation'' formulation}\\
As stated in section II, the method used to calculate $P(m,n)$ can be
viewed in terms of quantum-mechanical interacting particles. Here this
formulation is briefly reviewed and the recipe for calculating
quantities of interest is described.

The system studied in section II consists of a one-dimensional lattice
with sites numbered $1,2\cdots$, etc. A lattice site $k$, described by
the binary variable $\eta_k$, can either be occupied by a particle or a
vacancy. If a vacancy is present at site $k$ then $\eta_k=1$,
otherwise $\eta_k=0$. Hence, the state of a single system is described
by the set of variables $\eta\equiv\{\eta_k\}$ and can be represented as a
vector in a Fock space, i.e. 
\begin{eqnarray}
\ket{\eta}&=&\ket{\eta_1,\eta_2,\cdots\eta_k\cdots} \nonumber \\
\bra{\eta'}\eta\rangle&=&\prod_{k=1}^{\infty}\delta(\eta'_k,\eta_k)=\delta(\eta',\eta) \nonumber
\end{eqnarray}
Initially the system is full of $B$ particles, represented by the state $\ket{0}$. The
evolution of the system involves an $A$ particle being `shot' through
the $B$ array at time steps $m=1,2\cdots$, etc. Each $A$ particle passes
through a $B$ with probability $q$, or annihilates with the $B$ with
probability $p$. If no $B$ is present at a site the $A$ particle
passes through to the next site with probability one, see figure 1. After each time step $m$, the system will have some
probability $P_m(\eta)$ of being in the state $\eta$. The {\it vacancy} creation
and annihilation operators are now introduced
\begin{eqnarray}
C^+_k\ket{\cdots
0_k\cdots}&=&\left[p\prod_{i=1}^{k-1}q^{(1-\eta_i)}\right]\ket{\cdots 1_k\cdots}
\nonumber \\
C_k\ket{\cdots
1_k\cdots}&=&\left[p\prod_{i=1}^{k-1}q^{(1-\eta_i)}\right]^{-1}\ket{\cdots
0_k\cdots}
\nonumber 
\end{eqnarray}
with the auxiliary relations $C_k^+\ket{\cdots
1_k\cdots}=C_k\ket{\cdots 0_k\cdots}=0$. Defining the
$q$-commutator as $[A,B]_q=AB-qBA$, the following
commutations relations hold for $n_1<n_2$
\begin{eqnarray}
\left[C^+_{n_1},C^+_{n_2}\right]_q=\left[C_{n_2},C^+_{n_1}\right]_q=\left[C_{n_1},C^+_{n_2}\right]_q=\left[C_{n_2}^{},C_{n_1}^{}\right]_q=0& &\left[C_n,C^+_n\right]_+=1 \label{comm}
\end{eqnarray}
where the final anticommutator is for same-site operators.  These
commutation relations provide a representation of $SU_q(2)$, i.e. `$q$-deformed'
spin-half particles \cite{SU2Q}.  The evolution
equation for the system can now be written in terms of these vacancy creation and
annihilation operators 
\begin{eqnarray}
\ket{P_m}&=&\sum_\eta P_m(\eta)\ket{\eta}=\hat{T}^m\ket{0} \nonumber \\
\hat{T}&=&\sum_{k=1}^{\infty}C^+_k \nonumber
\end{eqnarray}
where $\hat{T}$ is the evolution operator. Introducing the left state $\bra{S}=\sum_{\eta'}\bra{\eta'}$ any observable $A(P_m)$ represented
by the operator $\hat{A}$ can be written in the following way 
\begin{eqnarray}
\bra{S}\hat{A}\hat{T}^m\ket{0}&=&\sum_{\eta'}\sum_{\eta}\bra{\eta'}\eta\rangle
A(\eta)P_m(\eta) \nonumber \\
\bra{S}\hat{A}\hat{T}^m\ket{0}&=&\sum_{\eta}A(\eta)P_m(\eta). \nonumber
\end{eqnarray}
In evaluating such observables it is useful to use a factorisation
characteristic of the creation operators. If a string of creation operators
is arranged, such that each operator acts in descending order (with
respect to its subscript) on the
right empty state, then the following property can be used
\begin{eqnarray}
\bra{S} C^+_{n_1}C^+_{n_2}\cdots
C^+_{n_m}\ket{0}&=&\bra{S}C^+_{n_1}\ket{0}\bra{S}C^+_{n_2}\ket{0}\cdots\bra{S}C^+_{n_m}\ket{0}
\label{fac}
\end{eqnarray}
where the single-vacancy expectations are simply $\bra{S}C^+_k\ket{0}=pq^{k-1}$. 

In the formalism described above, all quantities expressible in terms
of operators can be evaluated. Therefore
\begin{eqnarray}
\Psi(n_1,\cdots,n_m)&=& \bra{S}\sum_{perms}C^+_{n_1}C^+_{n_2} \cdots C^+_{n_m}\ket{0}
\nonumber\\
V_m(n)&=&\bra{S}C^+_n C_n\hat{T}^{m}\ket{0}
\nonumber \\
P(m,n)&=&\bra{S}C^+_n\hat{T}^{m-1}\ket{0}
\nonumber.
\end{eqnarray}
are the forms for each of the objects calculated in section II. These
can be evaluated by using the commutation relations (\ref{comm}) in the following
way. Firstly for a given string, commute all annihilation operators to
the right of the string. These objects when acting on the left state
$\bra{S}$ will leave strings containing only creation operators. These
strings can then be rearranged by using the appropriate relation in (\ref{comm}) so that they act on
the zero state in descending order of subscript. Finally the factorisation
property (\ref{fac}) can be used to calculate the expectation value. 

Interestingly, the algebraic structure outlined here is identical to
that used in the description of the partially-asymmetric exclusion
process (PASEP) with reflecting boundary conditions, therefore corresponding to a noisy Burgers equation with zero {\it average}
current  \cite{SaSch}. Microscopically, the PASEP
describes systems of particles that hop in a preferred direction with
a repulsive interaction. Assuming that the particles hop with a leftwards
bias, the density at site $n$ seen when $m$ such particles are
confined to a semi-infinite lattice ($[1:\infty]$), is identical to the vacancy density
$V_m(n)$ given by equation (\ref{shock1}). 

%
%

\newpage

\nsubsecnn{Figure captions}

FIGURE 1:

(i) A realisation of the model described in section I. The solid
lines are paths of
particles in space and time, with the dotted lines the trajectories annihilated particles would have taken. The $A$ and $B$ particles
start at positions $y_m$ and $z_n$ respectively and move with
fixed velocities $v_A=c$ and $v_B=-c$. An
intersecting pair of particles either annihilates (probability $p$)
or continues unaffected (probability $q=1-p$). 

(ii) The simplified, mapped system described in section II. At
unit time steps $A$ particles are `shot' through the $B$ lattice,
passing through each occupied site (with the same microscopic reaction
probabilities as above). A reaction occurring at a site produces a
vacancy, and it is the statistics of these vacancies that is used to derive
$P(m,n)$, the pair-reaction probability.

FIGURE 2:

(i) The steady-state density profiles of the $A$
and $B$ reactants at $x_r=s/\varrho$ for
the case $p=q=1/2$. The densities given in equation (\ref{profz})
involve the vacancy densities
(\ref{shock1}) for the exact case (circles and squares) and (\ref{shock2}) for the
$O(p)$ approximation (dashed lines).

(ii) The steady-state reaction rate at position $x_r=s/\varrho$, again
for  $p=q=1/2$. The the exact result (circles) is given in equation
(\ref{intrinsz}) with $x_r$ measured from the origin, or in equation (\ref{INT}) with $x_r$ measured from the centre of the fluctuating
overlap region. Also plotted are the mean-field (triangles) and the
$O(p)$ approximations (dashes) given in equations (\ref{steadymf}) and
(\ref{Papp2}) respectively. Comparison between the exact and mean-field results show the
reaction rate to be most correlated at the centre of the overlap region.

\newpage
\begin{figure}
\begin{minipage}[b]{8cm}
\epsfxsize 8 cm
\epsfysize 7 cm
\epsfbox{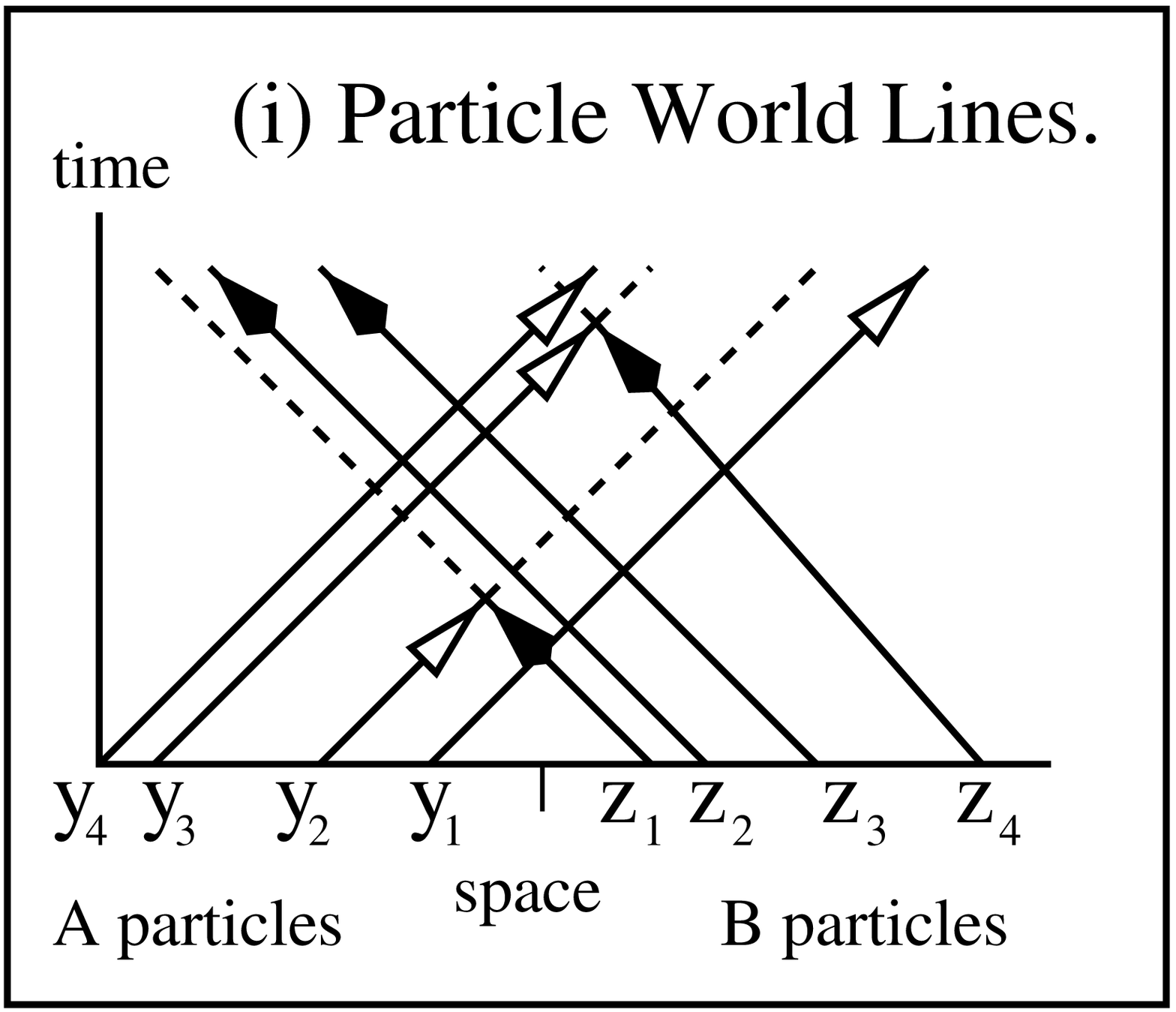}
\end{minipage}
\begin{minipage}[b]{8cm}
\epsfxsize 8 cm
\epsfysize 7 cm
\epsfbox{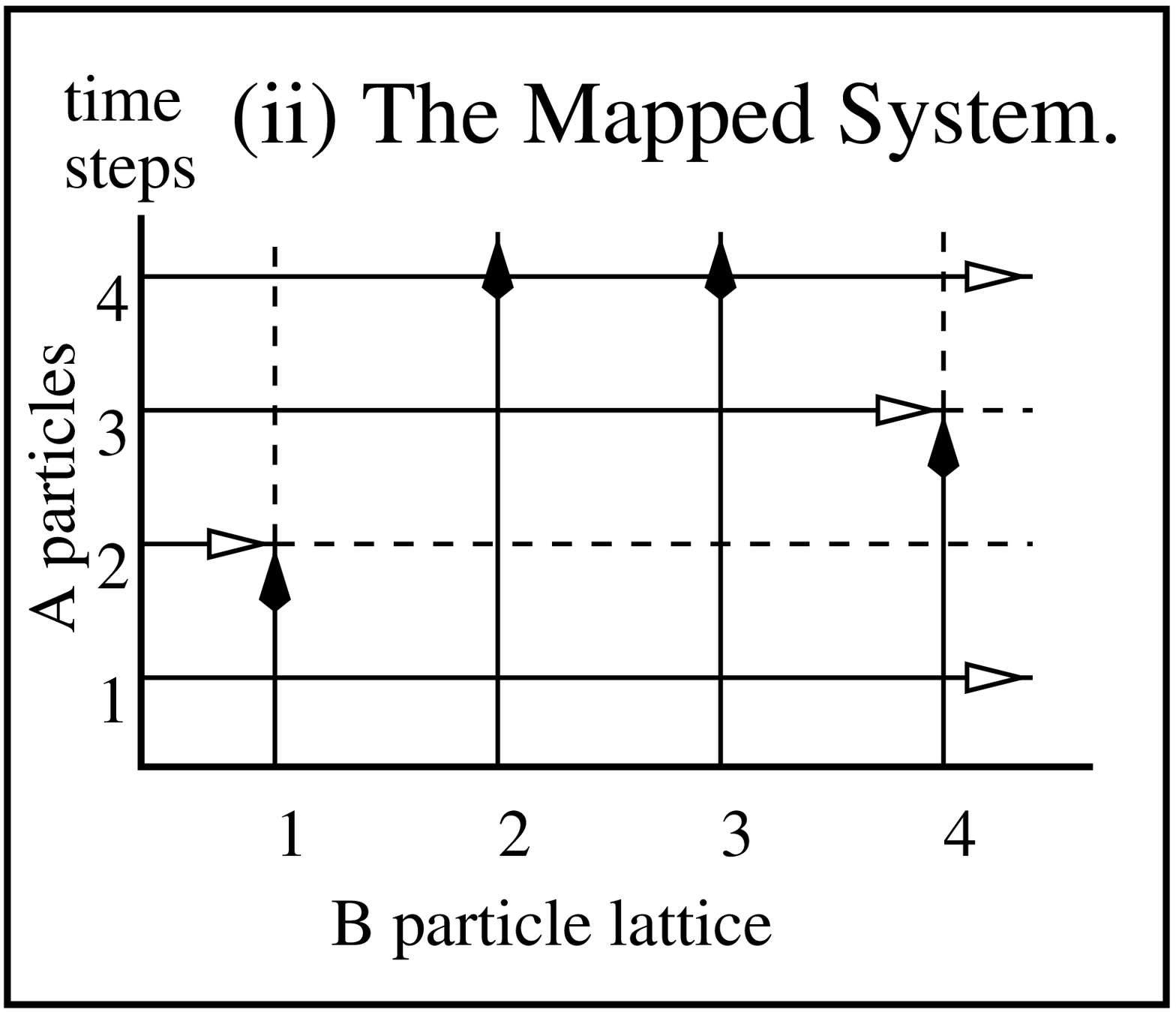}
\end{minipage}
\caption{}
\end{figure}

\begin{figure}
\begin{minipage}[b]{8cm}
\epsfxsize 8 cm
\epsfysize 7 cm
\epsfbox{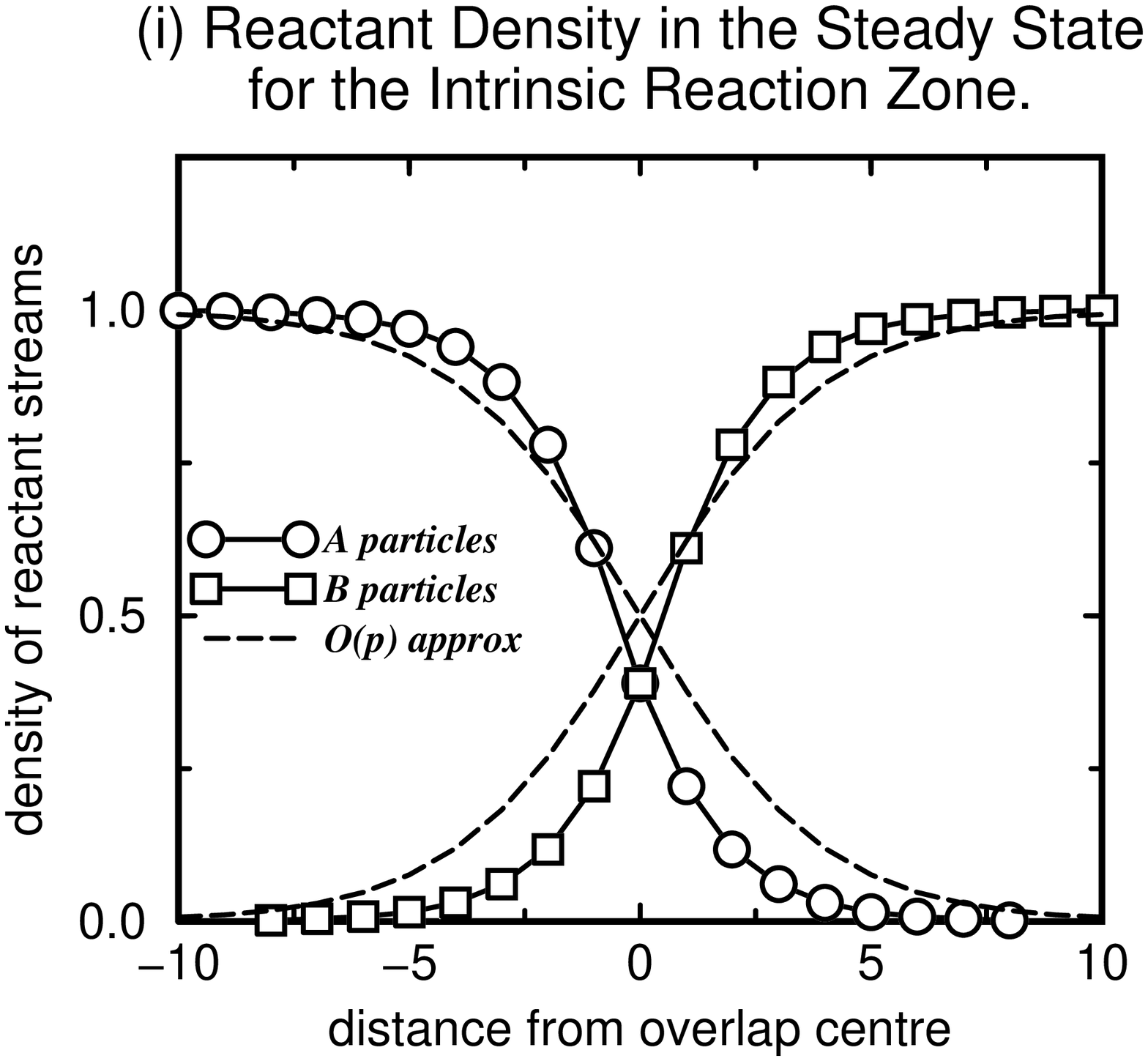}
\end{minipage}
\begin{minipage}[b]{8cm}
\epsfxsize 8 cm
\epsfysize 7 cm
\epsfbox{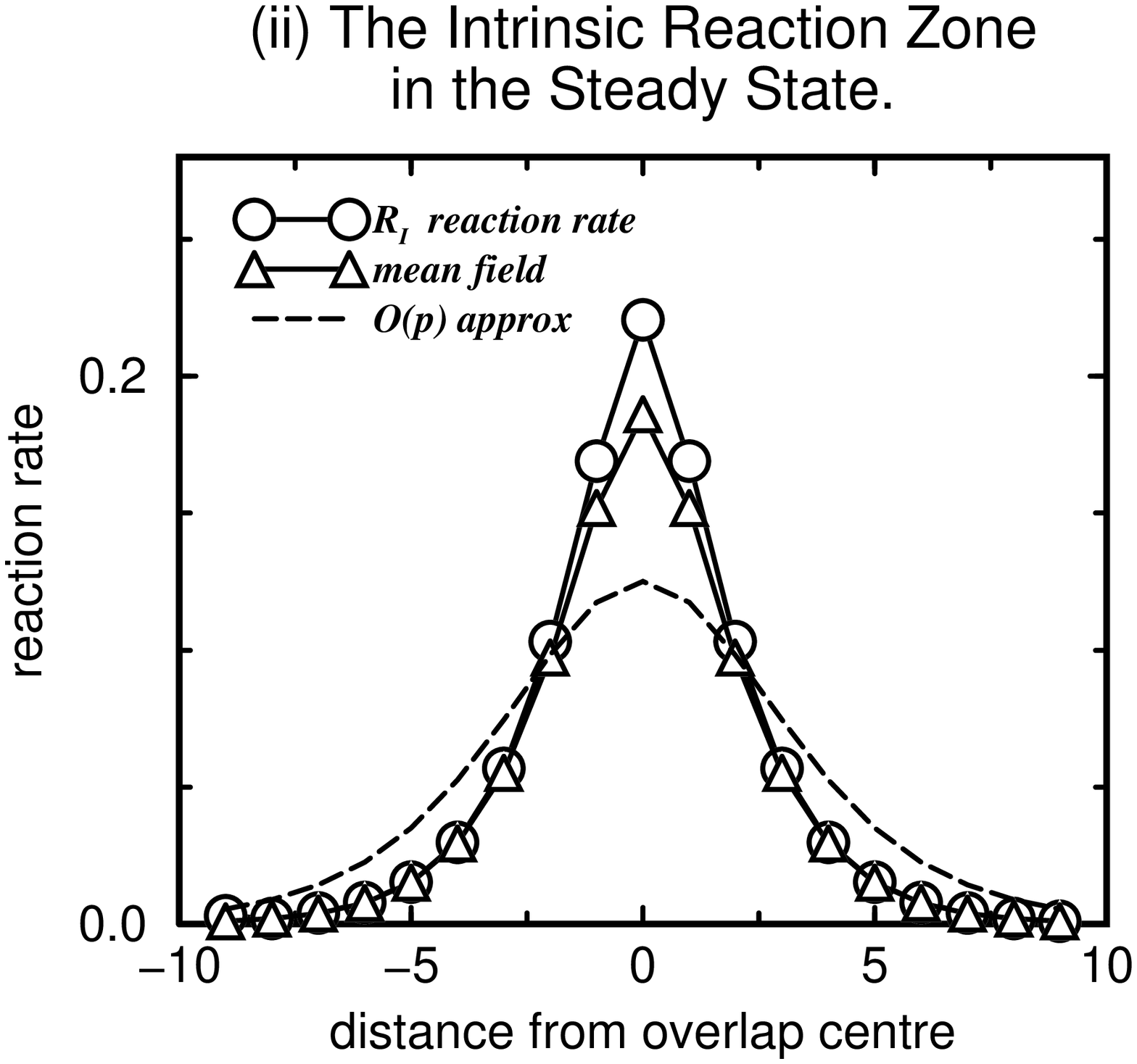}
\end{minipage}
\caption{}
\end{figure}

\begin{thebibliography}{99}
\bibitem{general}
	For a recent review see S. Redner in 
	{\it Nonequilibrium Statistical Mechanics in One
	Dimension}, ed V. Privman, Cambridge University Press 1997
\bibitem{ED} 
R. Kroon, H. Fleurent and R. Sprik  
{\it Phys. Rev. E} {\bf 47} (1993) 2462

\bibitem{MM} 
D. Toussaint and F. Wilczek 
{\it J. Chem. Phys.} {\bf 78} (1983) 2642

\bibitem{chem}
	Part VII - Experimental results.
	{\it Nonequilibrium Statistical Mechanics in One
	Dimension}, ed V. Privman, Cambridge University Press 1997
\bibitem{GR}
L. G\'alfi and Z. R\'acz 
{\it Phys. Rev. A} {\bf 38} (1988) 3151

\bibitem{EF}
Y. Elskens and H. L. Frisch
{\it Phys. Rev. A} {\bf 31} (1985) 3812

\bibitem{DRFP}
M. Droz, P. A. Rey, L. Frachebourg and J. Piasecki
{\it Phys. Rev. E} {\bf 51} (1995) 5541

\bibitem{Pias}
J. Piasecki
{\it Phys. Rev. E} {\bf 51} (1995) 5535

\bibitem{Droz1} 
J. Piasecki, P. A. Rey and M. Droz 
{\it Physica} {\bf 229A} (1996) 515

\bibitem{Droz2} 
P.-A. Rey, M. Droz and J. Piasecki 
{\it Euro. Jour. Phys.} (may 1997, to appear)

\bibitem{BR}
E. Ben-Naim and S. Redner 
{\it J. Phys A} {\bf 25} (1992) L575

\bibitem{BHC}
G. T. Barkema, M. J. Howard and J. L. Cardy 
{\it Phys. Rev. E} {\bf 53} (1996) R2017

\bibitem{BL}
M. Bramson and J. L. Lebowitz 
{\it J. Stat Phys} {\bf 65} (1991) 941

\bibitem{LC}
B. P. Lee and J. Cardy 
{\it Phys. Rev. E} {\bf 50} (1994) R3287

\bibitem{CORNDROZ}
S. Cornell and M. Droz 
{\it Phys. Rev. Lett.} {\bf 70} (1993) 3824 

\bibitem{RE}
M. J. E. Richardson and M. R. Evans 
{\it J. Phys. A} {\bf 30} (1997) 811

\bibitem{CDC}
S. Cornell, M. Droz and B. Chopard
{\it Phys. Rev. A} {\bf 44} (1991) 4826

\bibitem{AHLS}
M. Araujo, S. Havlin, H. Larralde and H. E. Stanley
{\it Phys. Rev. Lett.} {\bf 68} (1992) 1791

\bibitem{KRAPIVSKY}
P. L. Krapivsky
{\it Phys. Rev. E} {\bf 51} (1995) 4774

\bibitem{CORN}
S. Cornell 
{\it Phys. Rev. E.} {\bf 51} (1995) 4055

\bibitem{JANOWSKY}
S. A. Janowsky
{\it Phys. Rev. E.} {\bf 51} (1995) 1858

\bibitem{IKR}
I. Ispolatov, P. L. Krapivsky, and S. Redner
{\it Phys. Rev. E.} {\bf 52} (1995) 2540

\bibitem{turing}
A. Turing
{\it Philos. Trans. Roy. Soc. London Series B} {\bf 237} 37

\bibitem{SaSch}
S. Sandow and G. Sch\"utz 
{\it Europhys. Lett.} {\bf 26} (1994) 7

\bibitem{SANDOW}
S. Sandow 
{\it Phys. Rev. E} {\bf 40} (1994) 2660

\bibitem{FABVIV}
F. H. L. Essler and V. Rittenberg 
{\it J. Phys. A} {\bf 29} (1996) 3375

\bibitem{DERMEV}
B. Derrida, M. R. Evans, V. Hakim and V. Pasquier 
{\it J. Phys A} {\bf 26} (1993) 1493

\bibitem{SU2Q}
L. C. Biedenharn
{\it J. Phys. A} {\bf 22} (1989) L873

\bibitem{SHDOM}
G. Sch\"utz and E. Domany 
{\it J. Stat. Phys.} {\bf 72} (1993) 277

\end{thebibliography}
\end{document}